*Research Article*

# Complexity Measures for Quantifying Changes in Electroencephalogram in Alzheimer's Disease

**Ali H. Husseen Al-Nuaimi, Emmanuel Jammeh, Lingfen Sun, and Emmanuel Ifeachor**

*Signal Processing and Multimedia Communication (SPMC) Research Group, Faculty of Science and Engineering, School of Computing, Electronics, and Mathematics, University of Plymouth, Plymouth, UK*

Correspondence should be addressed to Ali H. Husseen Al-Nuaimi; ali.al-nuaimi@plymouth.ac.uk

Alzheimer's disease (AD) is a progressive disorder that affects cognitive brain functions and starts many years before its clinical manifestations. A biomarker that provides a quantitative measure of changes in the brain due to AD in the early stages would be useful for early diagnosis of AD, but this would involve dealing with large numbers of people because up to 50% of dementia sufferers do not receive formal diagnosis. Thus, there is a need for accurate, low-cost, and easy to use biomarkers that could be used to detect AD in its early stages. Potentially, electroencephalogram (EEG) based biomarkers can play a vital role in early diagnosis of AD as they can fulfill these needs. This is a cross-sectional study that aims to demonstrate the usefulness of EEG complexity measures in early AD diagnosis. We have focused on the three complexity methods which have shown the greatest promise in the detection of AD, Tsallis entropy (TsEn), Higuchi Fractal Dimension (HFD), and Lempel-Ziv complexity (LZC) methods. Unlike previous approaches, in this study, the complexity measures are derived from EEG frequency bands (instead of the entire EEG) as EEG activities have significant association with AD and this has led to enhanced performance. The results show that AD patients have significantly lower TsEn, HFD, and LZC values for specific EEG frequency bands and for specific EEG channels and that this information can be used to detect AD with a sensitivity and specificity of more than 90%.

## 1. Introduction

Alzheimer's disease (AD) is an age-related progressive, neurodegenerative disorder that is characterized by loss of memory and cognitive decline [1, 2] and it is the main cause of disability among older people [3]. AD is ranked as the sixth leading cause of death in US [4]. The rapid increase in the number of people living with AD and other forms of dementia due to the ageing population represents a major challenge to health and social care systems worldwide [5]. Currently, there are over 46.8 million individuals with dementia worldwide with an annual cost of care estimated at US$818 billion and is projected to reach 74.7 million by 2030 with an annual cost of US$2 trillion [6]. The number of individuals with dementia worldwide is expected to exceed 131 million by 2050 which will have a huge economic impact [7]. However, many dementia sufferers do not receive early diagnosis [7, 8]. It is estimated that up to 50% of people living with dementia may not have received formal diagnosis [8, 9]. In 2011, 28 million people of 36 million dementia sufferers did not receive a diagnosis worldwide [10].

Degeneration of brain cells due to AD starts many years before the clinical manifestations become clear [5, 11–15]. An early diagnosis of AD will contribute to the development of effective treatments that could slow, stop, or prevent significant cognitive decline [16–18]. An early diagnosis of AD could also be useful for identifying dementia sufferers who have not received a formal early diagnosis and this may provide an opportunity for them to access appropriate health care services [19–21].

A biomarker that can measure degeneration of brain cells due to AD at an early stage would be useful for its early diagnosis [2, 22–24]. But this may require dealing with large numbers of people as up to 50% of people living with dementia may not have received a formal diagnosis. Therefore, there is a need for simple, noninvasive, low-cost, and reliable biomarkers for early diagnosis which can be accessed in clinical practice [5, 25, 26]. Recent guidelines promote the use of biochemical and neuroimaging biomarkers to improve the diagnosis of AD. Cerebral spinal fluid (CSF) testing for AD is not widely used in clinical practice because it requires lumbar puncture which is an invasive procedure [2, 27, 28]. Neuroimaging is expensive, available only in specialist centres [29], and may not be suitable for patients that have pacemakers or certain implants [30]. Blood-based biomarkers have shown promising results in AD diagnosis but they are not yet fully developed and low-cost biosensors to detect AD biomarkers in blood do not exist at present [2, 25, 31].

Potentially, the electroencephalogram (EEG) can play a valuable role in the early diagnosis of AD [11, 20, 21, 24, 32–34]. EEG is noninvasive, low-cost, has a high temporal resolution, and provides valuable information about brain dynamics in AD [20, 21, 33, 35, 36]. The fundamental utility of EEG to detect brain signal changes even in the preclinical stage of the disease has been demonstrated [33, 37, 38]. Thus, EEG biomarkers may be used as a first-line decision-support tool in AD diagnosis [11, 35] and could complement other AD biomarkers [26].

AD is characterized by loss of memory and cognitive decline resulting from damage to brain cells which influence brain activity [38]. AD causes changes in the features of the EEG [35, 38, 39] and EEG analysis may provide valuable information about brain dynamics due to AD [20, 21, 33, 35]. The most characteristic features in EEG caused by AD are slowing of EEG, a decrease in EEG coherence, and reduction in EEG complexity [33–35, 37, 38, 40, 41]. These changes in the EEG can be quantified as a biomarker of AD. A variety of linear and nonlinear methods are being developed to quantify changes in EEG as AD biomarkers [42, 43]. AD biomarkers based on the slowing in EEG and a decrease in EEG coherence are often derived using linear analysis methods (i.e., spectral analysis of the EEG signal) [37, 44, 45], while biomarkers extracted by analysing the complexity of the EEG are based on nonlinear approaches (e.g., entropy methods, fractal dimension, and Lempel-Ziv complexity). The EEG complexity approaches have shown promising results in AD diagnosis [11, 35, 46] and appear to be appropriate for AD diagnosis [38, 47, 48]. Complexity is a measure of the extent to which the dynamic behavior of a given sequence resembles a random one [49]. The cortical areas of the brain fire spontaneously and this dynamic behavior of the brain is complex [50, 51]. AD causes a reduction in neuronal activity of the brain [52] resulting in decreased capability of the brain to process information [53–55] and this may be reflected in the EEG signals [52]. EEG complexity can potentially be a good biomarker for AD diagnosis [38] as AD patients have a significant reduction in EEG complexity [38, 40, 41, 52, 56, 57]. Several studies have investigated EEG complexity as a potential AD biomarker using whole EEG record with the objective of achieving a high performance. Given the association of EEG activities (e.g., alpha, delta activities) with AD, we hypothesized that the derivation of EEG complexity based on EEG activities should lead to enhanced performance.

This is a cross-sectional study aimed at demonstrating the usefulness of EEG based complexity measures to detect AD. In this study, we investigated an important class of complexity measures, information theoretic methods, which offers a potentially powerful approach for quantifying changes in the EEG due to AD [58]. Information theoretic methods (i.e., TsEn and LZC) have emerged as a potentially useful complexity-based approach to derive robust EEG biomarkers of AD [47, 58–62]. They are attractive because of the potential natural link between information theory-based biomarkers and changes in the brain caused by AD [58]. Conceptually, information processing activities in the brain are thought to be reflected in the information content of the EEG.

In particular, TsEn approach has been shown to be one of the most promising information theoretic methods for quantifying changes in the EEG [62, 63]. It has also been shown to be a reliable analysis tool to use with working memory tasks. As its computation is fast, it can serve as a basis for a real-time decision-support tool for dementia diagnosis by both specialists and nonspecialists [64]. Sneddon et al. [65] investigated TsEn of the EEG and was able to detect mild dementia due to AD with a sensitivity of 88% and specificity of 94%. De Bock et al. [62] found TsEn of the EEG to be a highly promising potential diagnostic tool for mild cognitive impairment (MCI) and early dementia with a sensitivity and specificity of 82% and 73%, respectively. Using TsEn approach, Al-Nuaimi et al. [35] detected AD from normal subjects with a sensitivity and specificity of 85.8% and 70.9%, respectively. Garn et al. [66] investigated the use of TsEn to diagnose AD based on EEG analysis and achieved a $p$ value < 0.0036 for channels T7 and T8 in discriminating between AD patients and normal subjects.

LZC is a nonparametric, nonlinear measure of complexity for finite length sequences [67]. It is a simple and powerful method which has been used in several biomedical applications [68]. LZC depends on a coarse-grain processing of the measurements [69] and can be applied directly on physiologic signal without preprocessing [70]. LZC has been applied extensively in analysing biomedical signals (e.g., EEG) to measure the complexity of discrete-time physiologic signals [67]. Furthermore, it is used to analyse brain function, brain information transmission, and EEG complexity in patients with AD [43]. The LZC approach produces a good biomarker for AD detection [70, 71]. Hornero et al. [72] used LZC to analyse EEG and magnetoencephalogram (MEG) in AD patients. They found that LZC provides a good insight into the EEG background activity characteristics and the changes associated with AD. Hornero et al. [73] found that LZC values were lower in AD patients and suggested that the most relevant differences are in the posterior region. In addition, they suggested that the MEG activity from AD patients is characterized by a lower degree of irregularity and complexity and that the LZC measures can be used to detect AD with a sensitivity and specificity values of 65% and 76.2%, respectively. McBride et al. [56] analysed EEG complexity based on the LZC method to discriminate between patients with early MCI, AD patients, and normal subjects. They found that EEG complexity features for specific EEG frequency bands with regional electrical activity provide promising results in discriminating between MCI, AD, and normal subjects. Fernandez et al. [74] analysed MEG complexity for MCI patients, AD patients, and normal subjects based on LZC method for discriminating between the three groups. They found that a combination of age and posterior LZC scores allowed them to distinguish between AD patients and MCI patients with 94.4% sensitivity and specificity.

HFD is a fast computational method for obtaining the fractal dimension of time series signals [75–77] even when very few data points are available [75]. It can track changes in a biosignal from a measure of its complexity [75, 76] and it is suited to capturing region-specific neural changes due to AD [45, 77]. In addition, HFD provides a more accurate measure of the complexity of signals compared to other methods [75, 78, 79] and it has been shown to be an efficient method for discriminating between AD patients and normal subjects [31, 80]. HFD of the EEG is potentially a good biomarker of AD diagnosis as it is significantly lower in AD patients than in normal subjects [46, 80, 81]. Smits et al. found that HFD is sensitive to neural changes selectively related to AD patients and normal subjects. Al-Nuaimi et al. [46] investigated HFD of EEG for AD diagnosis and they found that HFD is a promising EEG biomarker that captures changes in the regions of the brain thought to be affected first by AD and it could be used to detect AD with sensitivity and specificity values of 100% and 80%, respectively.

It is widely accepted that AD causes a decrease in the power of high frequencies (alpha, beta, and gamma) and an increase in the power of low frequencies (delta and theta) [11, 33, 34, 38, 41]. We hypothesized that complexity measures based on the EEG frequency bands would provide better results than those derived directly from the whole EEG record. The aim was to enhance the performance of the complexity measures and to demonstrate their usefulness in quantifying changes in EEG due to AD.

Digital filters were used to extract the five EEG frequency bands (i.e., delta, theta, alpha, beta, and gamma). Complexity measures were then obtained for each of the five EEG frequency bands and for each channel using each of the three methods of computing complexity measures (TsEn, HFD, and LZC).

For each method, we computed a panel of 114 biomarkers (i.e., 19 biomarkers for the whole EEG record and 19 biomarkers for each of the five EEG frequency bands). The performance measures for each biomarker were computed (including the sensitivity and specificity).

The paper is arranged as follows. In Section 2, the materials and methods used in the study are described. In Section 3, the results and discussions are presented and the conclusions are presented in Section 4.

## 2. Materials and Methods

*2.1. Materials.* This study was based on EEG dataset that was recorded from 52 volunteers. All the volunteers underwent a strict protocol based on normal hospital practices at Derriford Hospital, Plymouth, UK [11]. The EEG recordings include several states such as hyperventilation, awake, drowsy, and alert, with periods of eyes closed and open. For storage reasons, the sampling rate was reduced from 256 Hz to 128 Hz by averaging two consecutive samples. The duration of each EEG signal is 4 minutes. Figure 1 shows the electrode locations using a 10–20 system. The letters F, C, P, O, and T refer to cerebral cortex lobes (F: frontal, C: central, P: parietal, O: occipital, and T: temporal) [82].

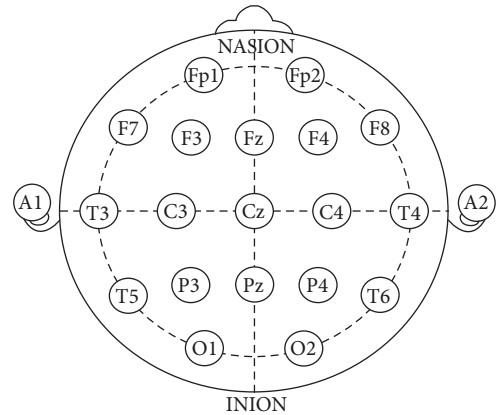

Figure 1: International 10–20 system.

The EEG dataset consists of two subdatasets (A and B). Subdataset A includes 11 age matched subjects over 65 years old (3 AD patients and 8 normal subjects). Subdataset A was recorded using the traditional 10–20 system in a Common Reference Montage by using the average of all channels as reference and the EEG signals were converted to Common Average and Bipolar Montages using software. Subdataset B includes 41 subjects that were not perfectly age matched (24 normal subjects, 10 males and 14 females, have mean age 69.4 ± 11.5 years (from 40 to 84 years) and 17 were probable AD patients, 9 male and 8 female). The normal subjects have a mean age of 69.4 ± 11.5 years (40 to 84 years) and the probable AD subjects have mean age of 77.6 ± 10.0 years (from 50 to 93 years). Subdataset B was recorded using the modified Maudsley system. The conventional 10–20 system has a similar setting with the Maudsley electrode positioning system [83].

All patients were referred to the EEG department at Derriford Hospital from a specialist memory clinic. A battery of psychometric tests (including the MMSE [84], Rey Auditory Verbal Learning Test [85], Benton Visual Retention Test [86], and memory recall tests [87]) were performed on all patients at the memory clinic. The classification of subjects with dementia was based on the working diagnosis provided by the specialist memory clinic. All healthy volunteers and AD patients had their EEG confirmed by a consultant clinical neurophysiologist at the hospital as normal and probable mild AD, respectively [11].

*2.2. Methods.* In our approach, the complete recordings of the EEG including artefacts were used without a priori selection of elements for analyses. This enabled us to have an idea about the robustness and usefulness of the method in practice. Data from a fixed interval (61 s to 240 s) was used to avoid electrical artefacts, which regularly occur at the beginning of a record, leaving a standard three-minute data to analyse.

The following steps outline the procedure that was used to derive the biomarkers for the three complexity methods (i.e., TsEn, HFD, and LZC)

(1) The EEG signal was filtered using infinite impulse response (IIR) Chebyshev-II bandpass filter into five frequency bands (i.e., delta 0–4 Hz, theta 4–8 Hz, alpha 8–12 Hz, beta 12–30 Hz, and gamma 30–45 Hz). A low computational IIR filter was used to retain the computational efficiency of the derived complexity-based biomarkers [88].

(2) The biomarkers were then derived first from the whole EEG record and then for each of EEG frequency bands for each of the three EEG complexity methods.

(3) For each biomarker of the EEG complexity methods (i.e., TsEn, HFD, and LZC), $p$ values were computed between AD patients and normal subjects using Student's $t$-test.

(4) The performance of each complexity measure to detect AD is then assessed. For each complexity measure, a classification model, based on the support vector machine (SVM), was used to detect AD.

*Tsallis Entropy (TsEn).* TsEn [89] biomarker computation of an $N$-samples EEG data sequence $x(1), x(2), \ldots, x(N)$ is based on the generalised measure of entropy, due to Tsallis:

$$\text{TsEn}_q = \frac{\left(\sum_{i=1}^{k} P_i - P_i^q\right)}{(q-1)}, \quad (1)$$

where $\text{TsEn}_q$ is the Tsallis entropy value, $k$ is the number of states that the amplitudes of the EEG are quantized into, $P_i$ is a probability associated with the $i$th state, and $q$ is Tsallis parameter ($k = 2200$ and $q = 0.5$).

*Higuchi Fractal Dimension (HFD).* To compute HFD biomarker [75, 77, 90] of an $N$-sample EEG data sequence $x(1), x(2), \ldots, x(N)$, the data is first divided into a $k$-length subdata set as

$$x_k^m: x(m), x(m+k), x(m+2k), \ldots,$$
$$x\left(m + \left[\frac{N-m}{k}\right] \cdot k\right), \quad (2)$$

where [ ] is Gauss' notation, $k$ is constant, and $m = 1, 2, \ldots, k$. The length $L_m(k)$ for each subdata set is then computed as

$$L_m(k)$$
$$= \left\{ \left[\sum_{i=1}^{[(N-m)/k]} |x(m+ik) - x(m+(i-1) \cdot k)|\right] \cdot ((N-1)/([(N-m)/k] \cdot k)) \right\} \cdot (k)^{-1}. \quad (3)$$

The mean of $L_m(k)$ is then computed to find the HFD for the data as

$$\text{HFD} = \frac{1}{K} \sum_{M=1}^{K} L_m(k). \quad (4)$$

*Lempel-Ziv Complexity (LZC).* To compute the LZC [43, 49, 67, 68, 70] biomarker of an $N$-sample EEG data sequence $x(1), x(2), \ldots, x(N)$, the EEG signal is first converted into a binary string as

$$x(i) = \begin{cases} 0 & \text{if EEG}(i) < M \\ 1 & \text{if EEG}(i) \geq M, \end{cases} \quad (5)$$

where $x(i)$ is the equivalent binary value of EEG$(i)$, $i$ is the index of all values in the EEG signal, and $M$ is the median value of each EEG channel. The median value is used to manage the outliers.

The binary string is then scanned from left to right until the end to produce new substrings. A complexity counter $c(N)$ is the number of new substrings. The upper bound of $c(N)$ is used to normalise $c(N)$ to get an independent value from the sequence of length $N$. The upper bound of $c(N)$ is $N/\log_2(N)$. $c(N)$ is then normalised by $b(N)$ as

$$C(N) = \frac{c(N)}{b(N)}, \quad (6)$$

where $C(N)$ is the normalised value of the LZC and $b(N)$ is the upper bound of the $c(N)$.

A panel of 114 biomarkers was computed (19 biomarkers for the whole EEG record and 19 biomarkers for each of EEG frequency band (i.e., delta, theta, alpha, beta, and gamma). To determine which features have a significant statistical association with AD, we computed $p$ values between AD patients and normal subjects using Student's $t$-test. This allowed us to identify significant features that may be useful to discriminate between AD patients and normal subjects. The dataset was split into training and testing data (60% for training and 40% for testing) with subjects selected at random. We selected 32 subjects for training and 20 subjects for testing at random from the datasets, a ratio of 60 : 40. The training data includes 12 AD (two from dataset A and 10 from dataset B) and 20 normal subjects (six from dataset A and 14 from dataset B). The testing data includes 8 AD (one from dataset A and seven from dataset B) and 12 normal subjects (two from dataset A and 10 from dataset B). $p$ values were computed using the training EEG dataset. Machine learning techniques were used to develop models based on the biomarkers. As a classifier, we used support vector machine (SVM) to model biomarkers extracted using TsEn, HFD, and LZC methods. SVM classifier was used because it is widely used in machine learning and has found application in dementia diagnosis. It has shown better performance in biomedical data analysis and in automatic AD diagnosis compared to other conventional classifiers (e.g., Euclidean distance classifier) and good capability to learn from experimental data [91, 92], and it has a stable classification performance [93].

It has also been shown to outperform other machine learning techniques (e.g., Naive Bayes, Multilayer Perceptron, Bayes Network, egging, Logistic Regression, and Random Forest,) in diagnosis of MCI and dementia [94]. We used the testing EEG dataset to test the performance of the models. For each complexity method, six performance tables were created (whole EEG record, and table for each EEG frequency band).

The performance of the TsEn, HFD, and LZC biomarkers for AD diagnosis was assessed in terms of sensitivity (Sen), specificity (Spec), accuracy (ACC), $F$-measure, error rate, true positive rate (TPR), false positive rate (FPR), positive predictive value (PPV), and negative predictive value (NPV). Matthew's correlation coefficient (MCC) was computed to measure the quality of the binary classification (AD and normal) between the actual and predicted results [95, 96].

## 3. Results and Discussions

*3.1. Result.* We analysed the performance of the three different complexity measures in quantifying changes in EEG due to AD. For this purpose, we examined the differences between the values of the complexity measures derived from EEG signals of AD subjects and those of normal subjects. Biomarkers that do not show significant differences between AD patients and normal subjects may not be suitable for quantifying changes in EEG due to AD as they may not be capable of being used to discriminate between AD patients and normal subjects.

We found that complexity measures derived from the EEG frequency bands for AD patients were significantly different to those of normal subjects compared to complexity measures derived from the whole EEG record. This suggests that they may be better suited to quantify changes in the EEG due to AD and potentially may provide better results in AD diagnosis.

Figure 2 shows the EEG biomarkers derived from whole EEG record (i.e., unfiltered) and those derived from the five EEG bands (delta, theta, alpha, beta, and gamma bands) using the TsEn method. The results show that TsEn values for AD patients are lower than those for normal subjects for whole EEG record. This is consistent with the findings in other studies [35, 58, 62, 64]. Figure 2 also shows that the differences between the TsEn values for AD patients and for normal subjects for the EEG bands (delta and theta bands in particular) are larger than those for whole EEG record. This is a desirable feature in a biomarker as it suggests that TsEn biomarkers derived from the EEG bands may provide better performance in detecting AD than those extracted from whole EEG record.

Figure 3 shows the EEG biomarkers derived from whole EEG record and those derived from the EEG bands (delta, theta, alpha, beta, and gamma bands in particular) using the HFD method. In this case, the results show that HFD values for AD patients are lower than those for normal subjects. This result is consistent with the finding in other studies [46, 80]. As with the TsEn, the differences between HFD biomarkers for AD patients and normal subjects for the EEG frequency bands (i.e., delta, theta, and alpha bands) were larger than those for the whole EEG record suggesting that the use of biomarkers derived from the frequency bands would be better at detecting AD than the use of whole EEG record.

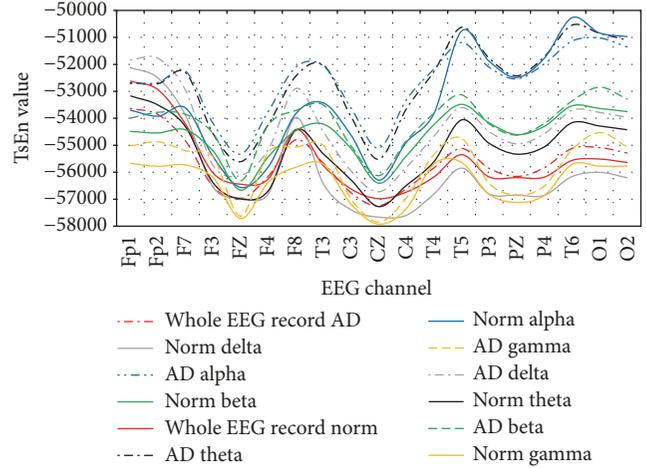

Figure 2: EEG biomarkers for TsEn.

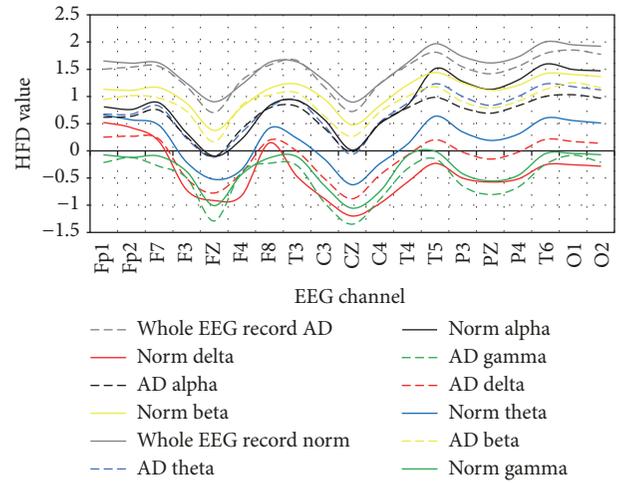

Figure 3: EEG biomarkers for HFD.

Figure 4 shows similar results for the LZC method. In this case, the results show that LZC values for AD patients were lower than those for normal subjects and these are consistent with the finding in other studies [43, 97]. Again, the differences between the LZC biomarkers for AD patients and normal subjects for the five EEG frequency bands (the theta, beta, and gamma bands, in particular) were larger those for the whole EEG record, suggesting that the use of biomarkers derived from the frequency bands would be better at detecting AD than the use of whole EEG record.

We analysed the complexity measures using $p$ values to determine the statistical significance in detecting AD

Figure 5 shows $p$ values of the differences in TsEn measures between AD patients and normal subjects for whole EEG record and those from the EEG frequency bands. The results show that TsEn biomarkers that were extracted from theta bands have the smallest $p$ values, while the TsEn biomarkers derived from gamma band have the maximum $p$ value between AD patients and normal subjects. This suggests that biomarkers that are extracted from theta band may provide the best performance in AD diagnosis. Figure 5 also shows that biomarkers that were extracted from EEG frequency bands may have a more significant association with AD than the EEG biomarkers that are derived from whole EEG record based on $p$ value analysis. Therefore, the complexity measures derived from the EEG frequency band may provide better results in the classification between AD patients and normal subjects.

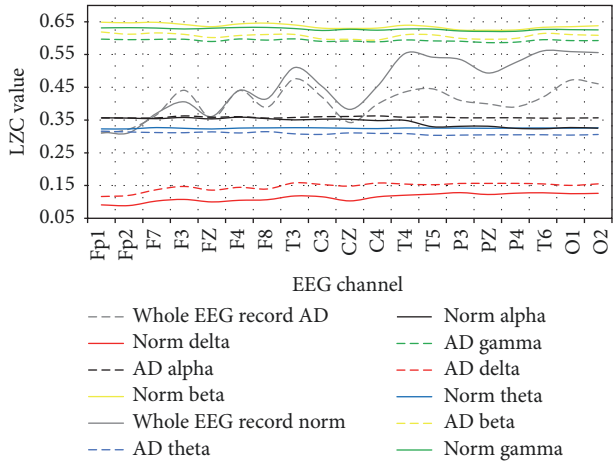

Figure 4: EEG biomarkers for LZC.

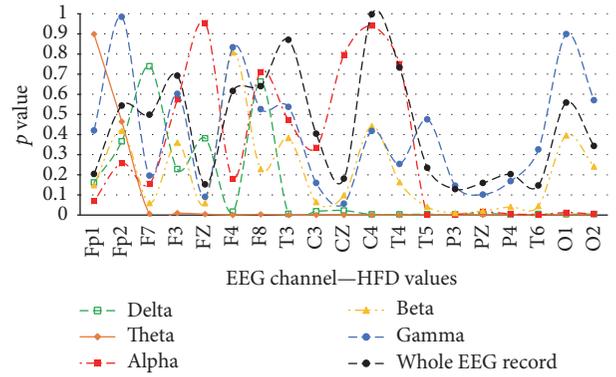

Figure 6: $p$ values for HFD between AD patients and normal subjects of the training EEG dataset.

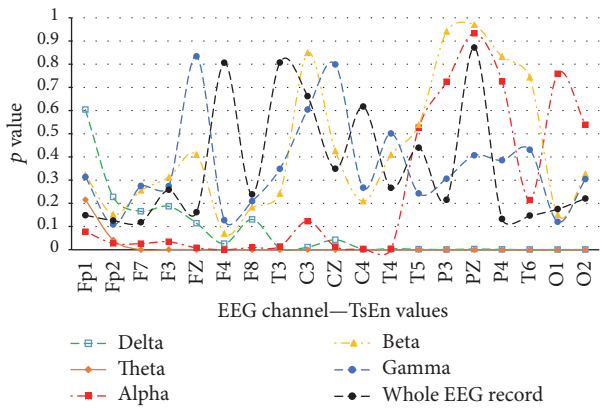

Figure 5: $p$ values for TsEn between AD patients and normal subjects of the training EEG dataset.

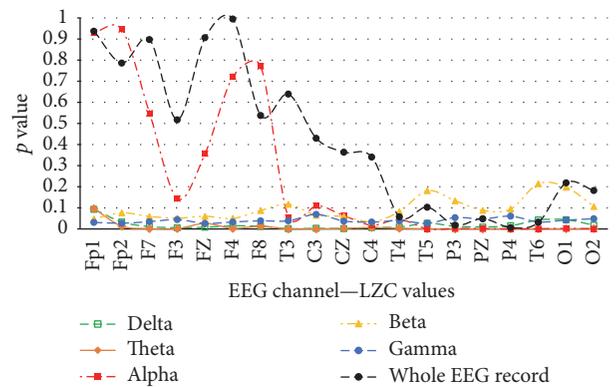

Figure 7: $p$ values for LZC between AD patients and normal subjects of the training EEG dataset.

Figures 6 and 7 depict the results of similar $p$ value analysis for HFD and LZC measures, respectively. The results show that, in both HFD and LZC methods, the complexity measures derived from the EEG frequency bands, theta band have significantly smaller $p$ values compared to those of measures derived from the whole EEG record. In both methods, complexity measures derived from the theta band gave the smallest $p$ value. This implies that biomarkers derived from the frequency bands, the theta band in particular, may provide the best possible performance in AD diagnosis using the HFD and LZC methods.

Looking across all the results (Figures 5, 6, and 7), the theta band has a minimum $p$ value between AD patients and normal subjects for all three complexity methods (i.e., TsEn, HFD, and LZC). Thus, EEG biomarkers derived from EEG frequency bands are better than the biomarkers that were extracted from whole EEG record. The biomarkers derived from theta band may provide the best performance in AD diagnosis across all three methods.

### 3.2. The Performance of the EEG Complexity-Based Measures.
Table 1 shows the performance of the SVM-based classification model using TsEn biomarkers for whole EEG record for the 19 EEG channels. In this case, the best sensitivity and specificity were 46.67% and 80%, respectively, for Fp2 and F7 EEG channels.

Similar performance indices were computed for each of the five EEG bands using the TsEn. As an example, Table 2 shows the performance indices for TsEn biomarkers for the delta band for the 19 EEG channels. The best sensitivity and specificity were 85.71% and 84.62%, respectively, for T4, O1, and O2 EEG channels.

Similar performance indices were computed for each of the five EEG bands using HFD and LZC methods. Table 3

ComplexityComplexity                                                                                           7

Table 1: TsEn performance for whole EEG record.

| EEG channel | Sen.% | Spec.% | Acc.% | F-measure% | Error rate | MCC | FPR% | FNR% | PPV% | NPV% |
|---|---|---|---|---|---|---|---|---|---|---|
| Fp1 | 43.75 | 75.00 | 50.00 | 58.33 | 0.50 | 0.153 | 25.00 | 56.25 | 87.50 | 25.00 |
| *Fp2* | *46.67* | *80.00* | *55.00* | *60.87* | *0.45* | *0.236* | *20.00* | *53.33* | *87.50* | *33.33* |
| *F7* | *46.67* | *80.00* | *55.00* | *60.87* | *0.45* | *0.236* | *20.00* | *53.33* | *87.50* | *33.33* |
| F3 | 43.75 | 75.00 | 50.00 | 58.33 | 0.50 | 0.153 | 25.00 | 56.25 | 87.50 | 25.00 |
| FZ | 44.44 | 100.00 | 50.00 | 61.54 | 0.50 | 0.272 | 0.00 | 55.56 | 100.00 | 16.67 |
| F4 | 44.44 | 100.00 | 50.00 | 61.54 | 0.50 | 0.272 | 0.00 | 55.56 | 100.00 | 16.67 |
| F8 | 44.44 | 100.00 | 50.00 | 61.54 | 0.50 | 0.272 | 0.00 | 55.56 | 100.00 | 16.67 |
| T3 | 37.50 | 50.00 | 40.00 | 50.00 | 0.60 | −0.102 | 50.00 | 62.50 | 75.00 | 16.67 |
| C3 | 35.71 | 50.00 | 40.00 | 45.45 | 0.60 | −0.134 | 50.00 | 64.29 | 62.50 | 25.00 |
| CZ | 42.11 | 100.00 | 45.00 | 59.26 | 0.55 | 0.187 | 0.00 | 57.89 | 100.00 | 8.33 |
| C4 | 44.44 | 100.00 | 50.00 | 61.54 | 0.50 | 0.272 | 0.00 | 55.56 | 100.00 | 16.67 |
| T4 | 35.29 | 33.33 | 35.00 | 48.00 | 0.65 | −0.229 | 66.67 | 64.71 | 75.00 | 8.33 |
| T5 | 33.33 | 50.00 | 40.00 | 40.00 | 0.60 | −0.167 | 50.00 | 66.67 | 50.00 | 33.33 |
| P3 | 28.57 | 33.33 | 30.00 | 36.36 | 0.70 | −0.356 | 66.67 | 71.43 | 50.00 | 16.67 |
| PZ | 37.50 | 50.00 | 40.00 | 50.00 | 0.60 | −0.102 | 50.00 | 62.50 | 75.00 | 16.67 |
| P4 | 35.71 | 50.00 | 40.00 | 45.45 | 0.60 | −0.134 | 50.00 | 64.29 | 62.50 | 25.00 |
| T6 | 26.67 | 20.00 | 25.00 | 34.78 | 0.75 | −0.471 | 80.00 | 73.33 | 50.00 | 8.33 |
| O1 | 27.27 | 44.44 | 35.00 | 31.58 | 0.65 | −0.287 | 55.56 | 72.73 | 37.50 | 33.33 |
| O2 | 30.00 | 50.00 | 40.00 | 33.33 | 0.60 | −0.204 | 50.00 | 70.00 | 37.50 | 41.67 |

Table 2: TsEn performance for delta band of the EEG signal.

| EEG channel | Sen.% | Spec.% | Acc.% | F-measure% | Error rate | MCC | FPR% | FNR% | PPV% | NPV% |
|---|---|---|---|---|---|---|---|---|---|---|
| Fp1 | 50.00 | 66.67 | 60.00 | 50.00 | 0.40 | 0.167 | 33.33 | 50.00 | 50.00 | 66.67 |
| Fp2 | 50.00 | 62.50 | 60.00 | 33.33 | 0.40 | 0.102 | 37.50 | 50.00 | 25.00 | 83.33 |
| F7 | 55.56 | 72.73 | 65.00 | 58.82 | 0.35 | 0.287 | 27.27 | 44.44 | 62.50 | 66.67 |
| F3 | 80.00 | 73.33 | 75.00 | 61.54 | 0.25 | 0.471 | 26.67 | 20.00 | 50.00 | 91.67 |
| FZ | 50.00 | 62.50 | 60.00 | 33.33 | 0.40 | 0.102 | 37.50 | 50.00 | 25.00 | 83.33 |
| F4 | 50.00 | 61.11 | 60.00 | 20.00 | 0.40 | 0.068 | 38.89 | 50.00 | 12.50 | 91.67 |
| F8 | 57.14 | 69.23 | 65.00 | 53.33 | 0.35 | 0.257 | 30.77 | 42.86 | 50.00 | 75.00 |
| T3 | 71.43 | 76.92 | 75.00 | 66.67 | 0.25 | 0.471 | 23.08 | 28.57 | 62.50 | 83.33 |
| C3 | 60.00 | 66.67 | 65.00 | 46.15 | 0.35 | 0.236 | 33.33 | 40.00 | 37.50 | 83.33 |
| CZ | 100.00 | 63.16 | 65.00 | 22.22 | 0.35 | 0.281 | 36.84 | 0.00 | 12.50 | 100.00 |
| C4 | 71.43 | 76.92 | 75.00 | 66.67 | 0.25 | 0.471 | 23.08 | 28.57 | 62.50 | 83.33 |
| *T4* | *85.71* | *84.62* | *85.00* | *80.00* | *0.15* | *0.685* | *15.38* | *14.29* | *75.00* | *91.67* |
| T5 | 80.00 | 73.33 | 75.00 | 61.54 | 0.25 | 0.471 | 26.67 | 20.00 | 50.00 | 91.67 |
| P3 | 75.00 | 83.33 | 80.00 | 75.00 | 0.20 | 0.583 | 16.67 | 25.00 | 75.00 | 83.33 |
| PZ | 100.00 | 75.00 | 80.00 | 66.67 | 0.20 | 0.612 | 25.00 | 0.00 | 50.00 | 100.00 |
| P4 | 83.33 | 78.57 | 80.00 | 71.43 | 0.20 | 0.579 | 21.43 | 16.67 | 62.50 | 91.67 |
| T6 | 83.33 | 78.57 | 80.00 | 71.43 | 0.20 | 0.579 | 21.43 | 16.67 | 62.50 | 91.67 |
| *O1* | *85.71* | *84.62* | *85.00* | *80.00* | *0.15* | *0.685* | *15.38* | *14.29* | *75.00* | *91.67* |
| *O2* | *85.71* | *84.62* | *85.00* | *80.00* | *0.15* | *0.685* | *15.38* | *14.29* | *75.00* | *91.67* |

summarises the best performance indices for the three complexity measures.

Figures 8, 9, and 10 summarise the performance indices of the TsEn, HFD, and ZLC methods.

The results show that TsEn, HFD, and ZLC EEG biomarkers derived from the EEG frequency bands provide better performance than EEG biomarkers derived from the whole EEG record.

3.3. Discussions. The results of this study show that EEG complexity-based measures provide a potentially useful way to detect AD. The most characteristic feature caused by AD is the reduction in EEG complexity [33–35, 37, 38, 40, 41] compared to normal subjects. This is consistent with other studies [35, 38, 43, 46, 56, 58, 62, 64, 80, 97, 98] and shows that EEG complexity measures are potentially a good biomarker for detecting AD.

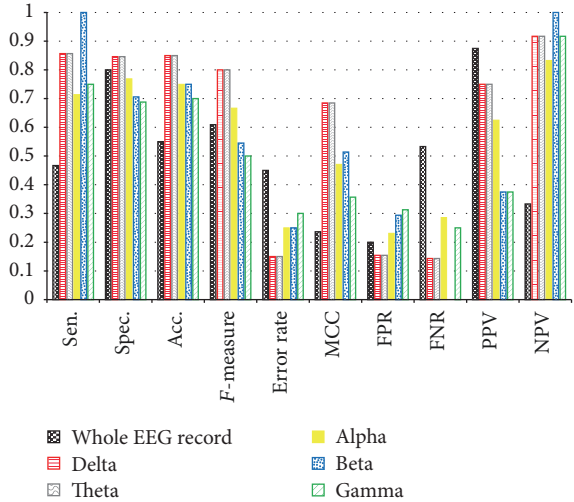

Figure 8: TsEn performance.

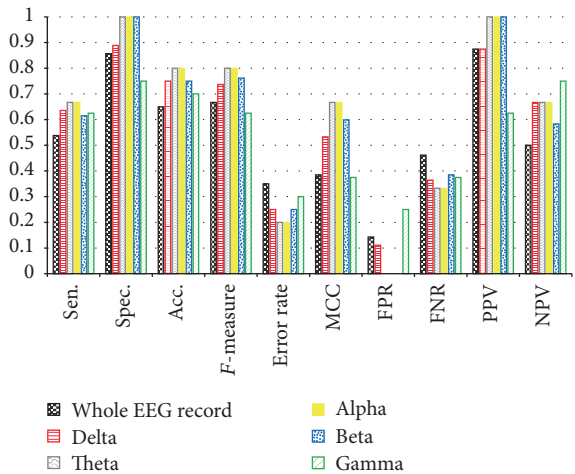

Figure 9: HFD performance.

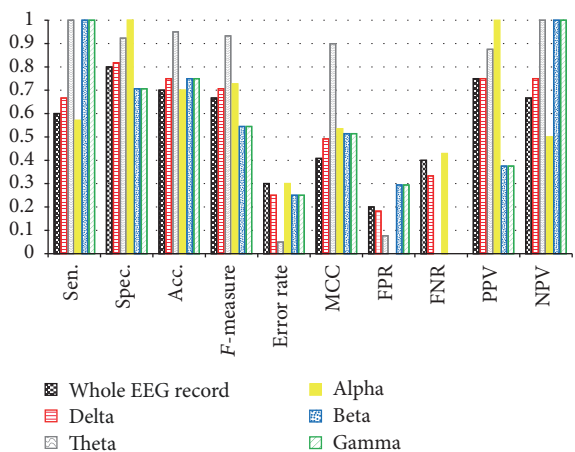

Figure 10: LZC performance.

Table 3: Summary of the best performance indices for the three complexity measures.

| Method | TsEn | | HFD | | LZC |
|---|---|---|---|---|---|
| Feature | Delta | Theta | Theta | Alpha | Theta |
| EEG channel | T4, O1, O2 | F4 | C4 | T5, P3 | C3 |
| Sen.% | 85.71 | 85.71 | 66.67 | 66.67 | 100 |
| Spec.% | 84.62 | 84.62 | 100 | 100 | 92.31 |
| Acc.% | 85 | 85 | 80 | 80 | 95 |
| F-measure% | 80 | 80 | 80 | 80 | 93.33 |
| Error rate | 0.15 | 0.15 | 0.2 | 0.2 | 0.05 |
| MCC | 0.685 | 0.685 | 0.667 | 0.667 | 0.9 |
| FPR% | 15.38 | 15.38 | 0 | 0 | 7.69 |
| FNR% | 14.29 | 14.29 | 33.33 | 33.33 | 0 |
| PPV% | 75 | 75 | 100 | 100 | 87.5 |
| NPV% | 91.67 | 91.67 | 66.67 | 66.67 | 100 |

Unlike previous studies, we found that the complexity measures derived from the EEG frequency bands (i.e., delta, theta, alpha, beta, and gamma) provide significantly better performance in detecting AD than the complexity measures derived from whole EEG records. This comes from the greater differences between the complexity measures for AD patients and normal subjects when they are derived from the frequency bands compared to when they are derived from whole record which is a desirable property of a good biomarker.

In particular, we found that for the TsEn and HFD complexity measures derived from the delta and theta bands gave the best performance. For the delta band, three EEG channels (T4, O1, and O2) gave the best performance. For the theta band, F4 gave the best performance.

Similar results were obtained for the LZC complexity measures, except that the best EEG channel was C3 for the theta band. This is consistent with the findings of other studies which suggested that AD starts from the back of the brain and then spreads gradually to other parts of the brain [5, 46, 99–101]. This implies that it may be possible to use only a small number of EEG channels to detect AD.

The findings of this study have a number of implications for research to develop new and robust techniques for the analysis of EEG to increase the contributions EEG makes to the diagnosis of AD.

The results suggest that the three EEG complexity measures, derived from the EEG frequency bands, can detect AD reliably (with sensitivity and specificity of >90%). Thus, EEG complexity measures could provide a basis for developing an accurate, low-cost, and easy to use tool to detect AD. Although the results of the studies are consistent with previous studies, unlike previous studies, in this study, the complexity measures are derived from EEG frequency bands (i.e., delta, theta, alpha, beta, and gamma). The results suggest that deriving the complexity measures from the EEG frequency bands is an important step for achieving robust biomarkers.

We found that AD patients have significantly lower complexity measures for specific EEG frequency bands and for specific EEG channels than normal subjects. This is consistent with findings in previous studies [33–35, 37, 38, 40, 41]. Thus, it may be possible to identify specific EEG channels and specific frequency bands that may provide the best biomarkers to detect AD. In situations where the number of available channels is limited (e.g., when portable EEG systems are used outside specialist centres), this may be exploited to achieve a good performance.

It may be possible to enhance the performance of the complexity-based approach further, by combining the three complexity measures into a composite model. Given that the three complexity measures are analysing different aspects of the signal (e.g., entropy and fractal measures), integrating them may lead to improved performance.

Our study has a number of limitations. At present, our methods have been applied only to the detection of AD, the most common form of dementia. A more detailed study is necessary to evaluate the methods using a much larger and diverse EEG datasets. This includes using the methods to differentiate between normal, MCI, and AD subjects [57, 63, 73].

This study shows that the abnormalities caused by AD can be detected by the complexity measures. However, similar changes may be caused by other neurodegenerative diseases, such as other types of dementia. To enhance the diagnostic usefulness of the methods, it may be necessary to develop them further to differentiate between dementias.

## 4. Conclusions

AD causes changes in the EEG due to loss of memory and cognitive decline and these changes are thought to be associated with functional disconnections among cortical areas resulting from the death of brain cells. Therefore, EEG analysis may provide valuable information about brain dynamics in AD. AD causes a reduction in neuronal activity of the brain and this may be reflected in EEG signals. Nonlinear methods based on EEG complexity approaches have shown promising results in detected changes in the EEG thought to be due to AD. Therefore, EEG complexity can potentially be a good biomarker for AD diagnosis. We investigated three complexity measures, TsEn, HFD, and LZC methods, derived from EEG frequency bands. We found that AD patients have significantly lower TsEn, HFD, and LZC values in specific EEG frequency bands and specific EEG channels compared to normal subjects. This may provide an effective way to discriminate between AD patients and normal subjects. Future work will evaluate the methods using larger and more diverse EEG datasets, including different types of dementia.

## Conflicts of Interest

The authors declare that there are no conflicts of interest regarding the publication of this article

## Acknowledgments

The first author would like to thank the Ministry of Higher Education and Scientific Research (MoHESR), Iraq, for their financial support. Financial support by the EPSRC is also gratefully acknowledged.

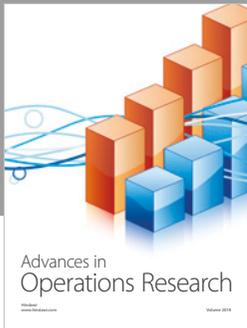 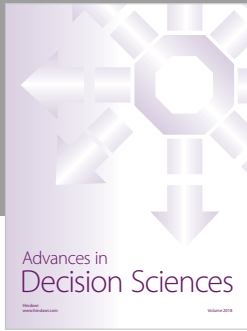 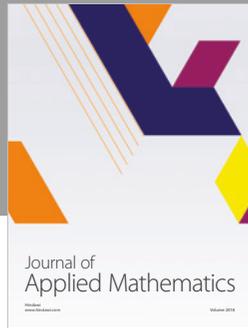 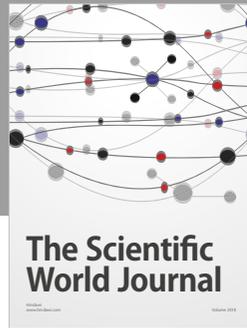 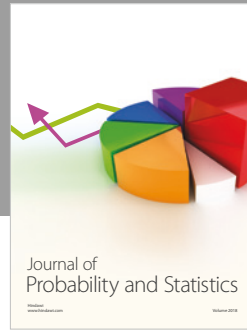
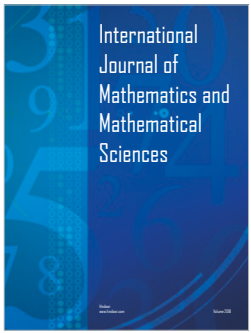 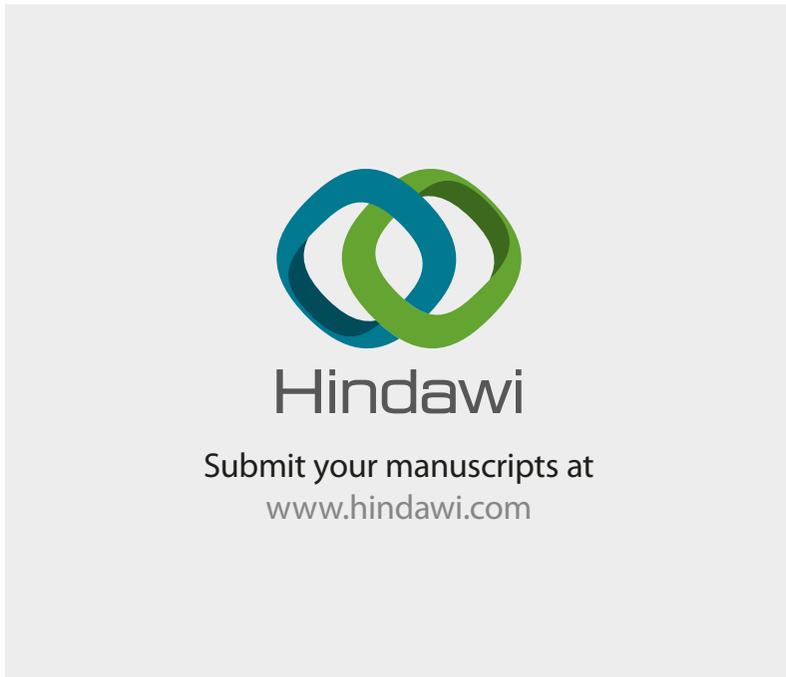 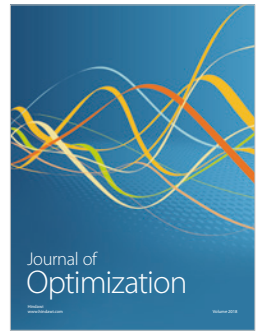
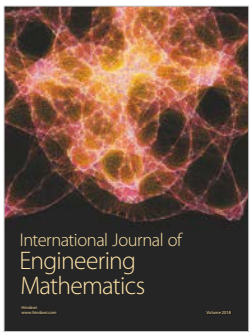 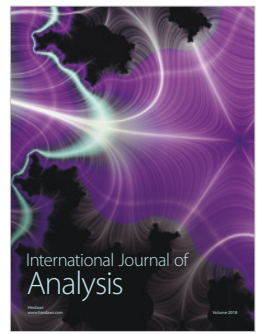
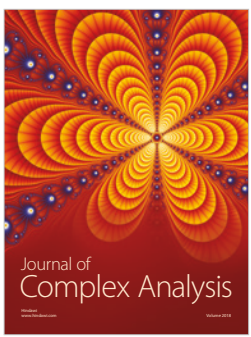 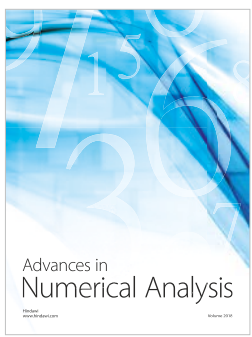 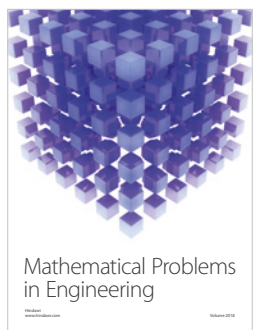 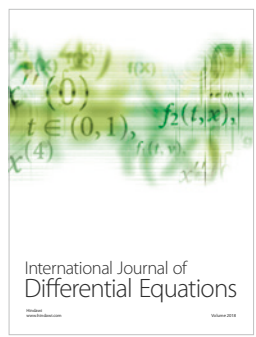 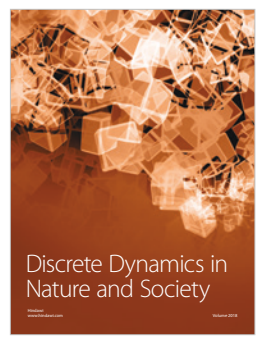
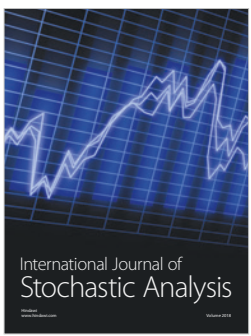 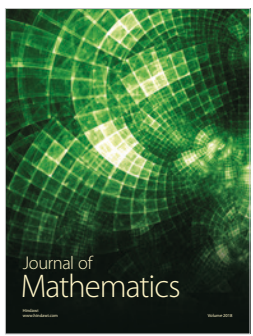 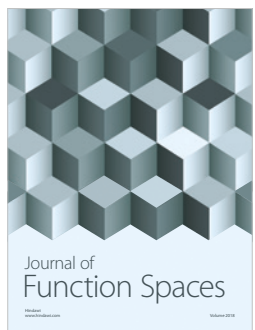 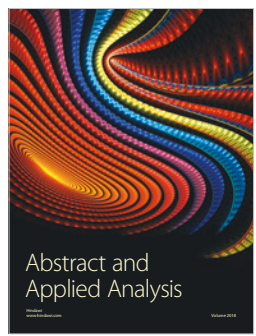 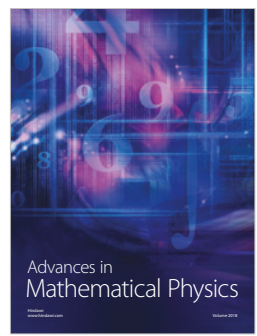

Submit your manuscripts at
www.hindawi.com